\documentclass[conference]{IEEEtran}
\usepackage{amsmath,amssymb}
\usepackage{array}
\usepackage{tabularx}
\usepackage[hidelinks]{hyperref}
\usepackage{graphicx}
\usepackage{placeins}
\usepackage{graphicx}

\begin{document}

\title{DICE: Tilting the Odds through Interface Attacks on System One Model}
\author{
\IEEEauthorblockN{
Ben Dong\IEEEauthorrefmark{1},
Tianyi Lu\IEEEauthorrefmark{2},
Zhonghao Guo\IEEEauthorrefmark{1},
Qian Wang\IEEEauthorrefmark{1}
}
\IEEEauthorblockA{
\IEEEauthorrefmark{1}University of California, Merced\\
Merced, California, USA  
}
\IEEEauthorblockA{
\IEEEauthorrefmark{2}Stevens Institute of Technology\\
Hoboken, New Jersey, USA
}
}

\maketitle

\begin{abstract}
System One models offer typed judgments that applications can turn
into decisions. Their answers, however, are conditional on the state
and question supplied by surrounding components, and on how the
application uses the response. We study this boundary in a Jev-based
pipeline with bounded substitutions at three adapter outputs: state
extraction, question generation, and output interpretation. Across
three small synthetic studies, each adapter can change the final
decision while Jev and its authentic API responses remain outside
attacker control. The cases expose distinct failures of source
fidelity, answer meaning, and response relevance. Checks that bind
these relationships stopped the corresponding interventions we
tested, although source verification also rejected some valid clean
decisions. The pilot identifies where correct request-level judgments
can lose their connection to the original task; it does not estimate
the prevalence of these failures in deployed systems.
\end{abstract}


\section{Introduction}

System One models expose a useful division of labor in decision
systems. An application supplies state and a question; the model
returns a typed assessment; application code decides what action
follows~\cite{deusser2026jev}. Jev is one such black-box service~\cite{deusser2026jev,li2026fast}. Its response can be
authentic and appropriate to the request while the application still
makes the wrong decision~\cite{saltzer1984endtoend}. The reason is that the request and its
eventual use are constructed outside the model~\cite{deusser2026jev,li2026fast}.

Consider a credential record that no longer permits an operator to
use a machine. A state-extraction adapter may advance the credential's
expiry in structured data. A question-generation adapter may preserve
the credential rules but ask whether permission should be withheld,
while downstream code continues to treat \emph{yes} as approval. An
output-interpretation adapter may pass along an authentic approval
answer from before the record was updated. Jev can answer each
submitted request correctly; none of those answers alone establishes
that the current operator should be approved. The three changes act
at different points, but each severs a relationship between the
original task and the final action.

We study those relationships as three attack pillars: \emph{state}
fidelity to the source, \emph{question} fidelity to the intended
decision, and \emph{output} fidelity to the current request. Our
adversary controls the output of exactly one adjacent LLM adapter at
a time. The source record, policy, final decision rule, Jev service,
and authentic request--response log remain protected. This setting
does not explain how an adapter becomes compromised. It isolates what
bounded control at one interface can do after compromise.

The distinction we test is between local reasoning correctness and
end-to-end decision correctness~\cite{saltzer1984endtoend,li2026fast,liu2026security}. We construct paired clean and
attacked traces with independently specified task decisions, then
check both Jev's answer to the actual request and the action selected
from that answer. The three synthetic studies exhibit wrong final
decisions from altered facts, reversed answer meaning, and reuse of
an earlier authentic response. The evidence is strongest as a
demonstration of these mechanisms, not as an estimate of how often
they occur in deployed applications. A separate cross-pillar set
varies rule structure, and exact-request repeats test response
stability without treating repeated calls as new tasks.

The defenses follow the same division. A source-aware check must
compare extracted facts with the protected record; a question check
must preserve what a positive answer means to the selector; and an
output check must bind evidence to the current request. In our pilot,
the corresponding checks stopped the tested interventions, while
state verification also rejected some faithful clean extractions.
These observations sharpen the research question: when does a
trusted model's valid judgment cease to support the application
decision, and what context must an auditor retain to recognize that
gap? We answer it for three bounded adapter manipulations around
Jev, without attributing the resulting decisions to a defect in the
model itself.

The Main contribution of this paper includes:
\begin{itemize}
    \item \textbf{A three-boundary account of decision integrity.}
    We distinguish state fidelity, question-to-action meaning, and
    response relevance as separate obligations around a trusted
    typed-judgment service.

    \item \textbf{Paired evidence from Jev-based decisions.}
    Across three synthetic studies, bounded single-adapter changes
    alter final decisions while authentic Jev responses and
    request-level correctness are assessed separately.

    \item \textbf{An evaluation of boundary checks.}
    We compare the context available to different checks, show
    which tested interventions they stop, and record clean-decision
    costs and limits of the observed results.
\end{itemize}

\section{Background and Related Work}

\subsection{Decision Model}

Jev is a commercial decision model developed by TypeSafe AI
and positioned as a ``System One'' component for software
automation. Given a shared state and typed questions, it
returns discrete choices, rubric-based scores, or boolean
probabilities, supporting classification, routing, assessment,
and automated verification; multiple questions can be evaluated
within a single request~\cite{vercel2026jev}.
TypeSafe AI describes its training approach as Reinforcement
Learning for Calibrated Decisions (RLCD), emphasizing typed
outputs and confidence estimates that software can use to
determine when to act or request review~\cite{typesafe2026systemone}.
Empirical evidence provides a more qualified picture:
Deu{\ss}er et al.\ evaluate Jev across 37 datasets and report
strong zero-shot performance and well-calibrated choice
probabilities, while finding that boolean outputs can require
task-specific threshold tuning and that fine-grained
classification and rubric-based judgments remain
challenging~\cite{deusser2026jev}.
Relevant research includes pretrained language understanding
models such as BERT~\cite{devlin2019bert}, rubric-based evaluation
with G-Eval~\cite{liu2023geval}, specialized evaluators such as
JudgeLM~\cite{zhu2023judgelm}, and learned routing with
RouteLLM~\cite{ong2025routellm}.
These works provide context for Jev's role as a reusable decision
component, while Jev-Mem demonstrates its application to
controlling memory organization and retrieval in AI
agents~\cite{jiang2026jevmem}.

OpenAI’s Decisions API evaluates text and images and returns typed answers for classification, routing, and prioritization. It supports three question types: predicates estimating whether a condition is true, choices selecting among predefined options, and scores evaluating inputs against ordered criteria. This interface connects model assessments to application decisions through probability estimates and application-defined thresholds\cite{openai_decisions}. 

System One and LLMs describe different dimensions: System 1 refers to fast, intuitive processing, whereas LLM refers to a model class. An LLM can produce rapid responses or support more deliberate reasoning. Decision Transformer connects sequence modeling with action prediction, while ReAct combines language-based reasoning with environmental interaction~\cite{yao2023react}. These works motivate studying how JEV’s fast decisions could be complemented by deliberate reasoning when uncertainty or task complexity increases.

\subsection{Prompt Attacks on LLM}

Prompt attacks seek to redirect LLM behavior through adversarial
instructions or crafted inputs. Prompt injection attempts to make a
model follow attacker-controlled instructions, including those embedded
in retrieved documents or tool outputs. Greshake et al.\ demonstrate
that such indirect injections can compromise LLM-integrated applications
by crossing the boundary between external data and trusted
instructions~\cite{IndirectPromptInjection10.1145/3605764.3623985}.
This exposes a broader application-level vulnerability: content intended
to provide evidence may instead influence how the application processes
a request or invokes its tools. Zhan et al.\ further investigate this
problem through InjecAgent, a benchmark of indirect prompt injections
against tool-integrated agents, covering attacks intended to harm users
or exfiltrate private data~\cite{zhan2024injecagent}.
Jailbreak attacks instead seek to bypass safety constraints; Zou et al.\
show that optimized adversarial suffixes can transfer across multiple
models~\cite{zou2023universal}. Although prompt injection and jailbreaking
have different objectives, both examine whether adversarial inputs can
override intended behavioral constraints.

These studies motivate examining the integrity of LLM adapters
surrounding Jev. In a pipeline where an adapter converts natural-language
requests and external evidence into Jev's state and typed questions,
an injection could alter extracted facts, omit relevant qualifications,
or substitute an attacker-selected question. A downstream adapter could
also misrepresent the returned decision when producing a user-facing
response. Such failures could yield an incorrect application outcome
even if Jev correctly evaluates the state and questions it receives.
Typed interfaces constrain output structure, but structural validity
alone does not establish that the submitted content faithfully
represents the original request~\cite{singh2026structured}.

Wallace et al.\ propose an instruction hierarchy that trains models to
prioritize privileged instructions and selectively ignore conflicting
instructions from lower-trust sources~\cite{wallace2024instructionhierarchy}.
This provides a relevant framework for studying adapter trust boundaries.
Accordingly, evaluating a Jev-based pipeline should examine whether
adversarial content changes the submitted state, questions, or final
interpretation, alongside the correctness of the resulting decision. 

\subsection{Related Research} 

Recent studies evaluate Jev as a security decision component.
Liu compares Jev, Laya, Decider, and Bespoke Nimble on
prompt-injection detection, interaction-risk judgment, and
harmful-request screening, examining accuracy, calibration,
and selective automation~\cite{liu2026security}.
The study shows that aggregate accuracy and calibration can
conceal high-confidence security errors, and recommends
validating question definitions, probability thresholds,
and policy actions separately. Li conducts a paired evaluation
of Jev and Laya across 11 agent decision points, including
injection screening and tool selection~\cite{li2026fast}.
Its evaluation and self-audit demonstrate the need to distinguish
component accuracy from end-to-end quality and deployment cost.
These studies motivate assessing decision models in their
application context. Our work complements them by examining
whether bounded manipulation of a surrounding adapter can
produce an incorrect final decision while Jev correctly answers
the request it actually receives.

Research on agent security examines related risks across
tool use and external information processing. AgentDojo provides
an extensible environment for evaluating prompt-injection
attacks and defenses against agents operating over untrusted
tool outputs~\cite{debenedetti2024agentdojo}.
Agent Security Bench (ASB) evaluates attacks and defenses
across prompts, tools, and memory, including memory poisoning
and backdoor attacks~\cite{zhang2025asb}.
ToolSafe focuses on step-level tool invocation safety:
TS-Guard assesses proposed actions using interaction history
before execution, while TS-Flow uses guardrail feedback
to guide subsequent agent behavior~\cite{mou2026toolsafe}.
Dong et al.~\cite{dong2026claws} examine unauthorized context
disclosure through LLM-generated tool-call arguments. Their
Claw in Plain Sight attack uses task-adjacent authority pressure
to induce the inclusion of protected attributes in otherwise
valid arguments. Their synthetic evaluation measures
policy-violating argument generation rather than completed
network exfiltration, motivating purpose- and destination-aware
checks before tool execution. This work studies privacy at the
context-to-argument boundary; DICE studies decision integrity
across state extraction, question generation, and response
interpretation around a trusted decision service. These approaches connect security assessment to agent behavior
and legitimate task completion. Our studies instead isolate
three interfaces surrounding a trusted decision service:
state extraction, question generation, and response
interpretation. We assume control of one adapter rather than
demonstrating how an attacker obtains that control.

Several defenses address the boundaries between trusted
instructions, untrusted data, and executable actions.
StruQ separates instructions and data through a secure front
end and a specially trained model, while SecAlign uses
preference optimization to favor legitimate instructions
over injected ones~\cite{chen2025struq,chen2025secalign}.
CaMeL separates control and data flows and applies
capability-based policies to constrain tool
use~\cite{debenedetti2025camel}.
These works establish the importance of protecting the
architecture surrounding language models. DICE investigates
three specific integrity failures within a Jev-based pipeline:
an extracted state that differs from the protected source,
a question whose positive answer has the wrong action meaning,
and an authentic response associated with an earlier request.
Accordingly, we evaluate checks that bind extracted facts
to source inputs, answer semantics to the intended action,
and interpreted responses to the current request.
Our contribution is an empirical demonstration of these
interface mismatches under bounded single-adapter control,
with request-level correctness assessed separately from
final decision integrity.

\section{Threat Model}
\label{sec:threat}

\subsection{System Model}

\begin{figure*}[t]
    \centering
    \includegraphics[width=\textwidth]{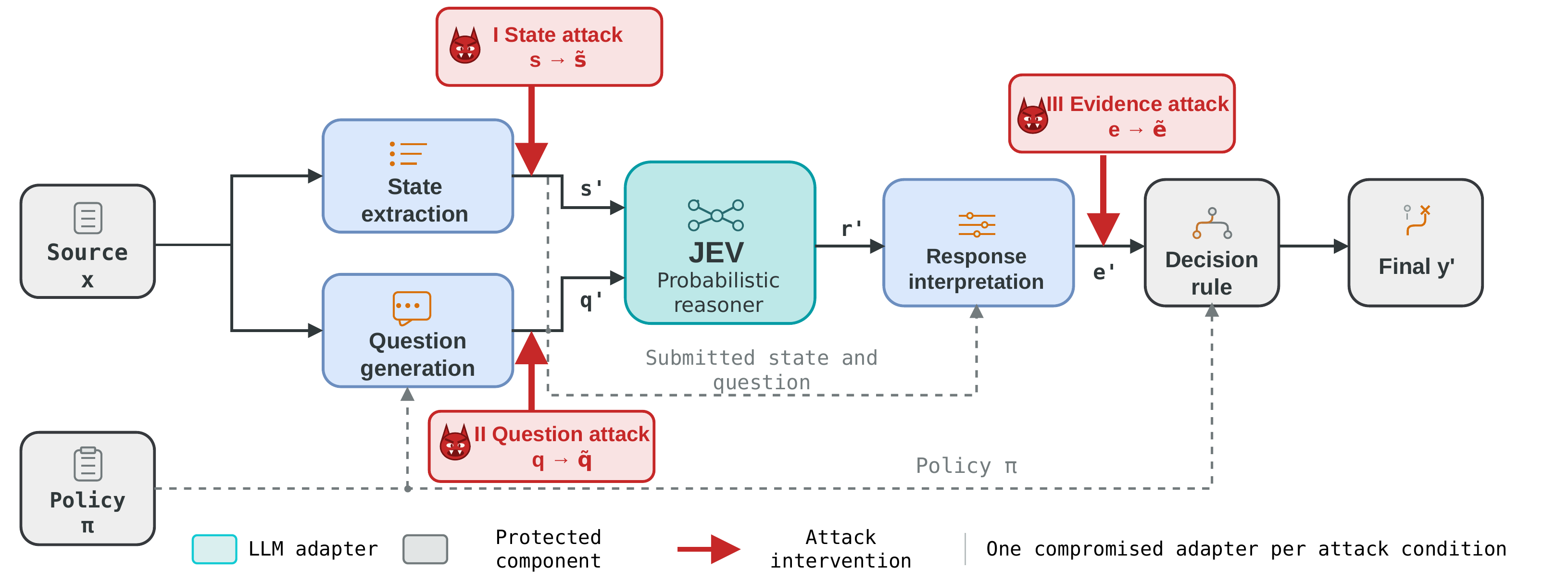}
    \caption{Jev decision pipeline and interface attacks.
    Red arrows indicate interventions on structured state,
    submitted questions, and interpreted evidence.
    Each attack condition compromises one LLM adapter;
    Jev and the final decision rule remain protected.}
    \label{fig:interface-attacks}
\end{figure*}

We consider a decision pipeline in which upstream large language model
(LLM) adapters construct the input to a trusted reasoning service and a
downstream LLM adapter interprets its response.
As illustrated in Figure~\ref{fig:interface-attacks}, the three adapter
roles are state extraction, question generation, and output
interpretation. Jev serves as the trusted black-box reasoning
component, accessible only through its API. The intended decision
policy guides question generation and the final decision rule,
while the interpretation adapter receives the submitted state and
question together with Jev's response. The pipeline is represented as
\begin{align}
 s &= E(x), & q &= Q(s,\pi), \nonumber\\
 r &\sim J(\cdot\mid s,q), & e &= I(s,q,r), \label{eq:pipeline}\\
 y &= F(e,\pi). && \nonumber
\end{align}
Here $x$ is the source, $\pi$ the protected policy, $r$ the
authentic Jev response, and $F$ the protected decision rule.
Figure~\ref{fig:interface-attacks} uses primes for values in a
compared execution: $s'$, $q'$, $r'$, $e'$, and $y'$. A tilde marks
only the value directly substituted at the compromised adapter,
such as $\widetilde{s}$, $\widetilde{q}$, or $\widetilde{e}$.
Other primed values may change as downstream consequences of that
single substitution. The response $r'$ remains an authentic Jev
answer to the submitted pair $(s',q')$.

Figure~\ref{fig:interface-attacks} identifies three intervention points.
A \emph{state attack} replaces the extracted state $s$ with
$\tilde{s}$, potentially changing the facts presented to Jev.
A \emph{question attack} replaces the generated question $q$ with
$\tilde{q}$, potentially redirecting the evaluation away from the
intended policy.
An \emph{evidence attack} replaces the interpreted evidence $e$ with
$\tilde{e}$ before it reaches the final decision rule, potentially
misrepresenting Jev's authentic response.
We evaluate these attack conditions separately, with only one adapter
compromised in each condition. The raw source, intended policy,
Jev service, and final decision rule remain protected.

The primed quantities $s'$, $q'$, $r'$, $e'$, and $y'$ in
Figure~\ref{fig:interface-attacks} denote the values realized along
the attacked pipeline. They include both the directly manipulated
quantity and any downstream values affected by that intervention.
In particular, $r'$ remains an authentic response from Jev to the
submitted inputs, even when those inputs have been manipulated.
This distinction separates reasoning-service correctness from
end-to-end decision integrity: a correct response to an altered
state or question, or an incorrect interpretation of an authentic
response, can still produce a final decision that violates the
intended policy.

\subsection{Trusted Components and Correctness}
The reasoning service, original input, intended policy, final decision
rule, and recording of actual API requests and responses are outside
attacker control. The service's parameters and implementation are
unavailable to the attacker. Trust in the service means that it is
uncompromised; correctness on a particular request is a separate
property to be checked.

Let $y^*(x,\pi)$ denote the independently specified correct task decision.
Let $C(s,q,r)$ indicate whether the response correctly answers the
question actually submitted under the submitted state. These notions
are distinct: a correct answer to an altered question or altered state
need not support the correct decision for the original task. Response
authenticity alone does not establish either form of correctness.

\subsection{Attacker Objective and Audit Visibility}
The objective is an incorrect final decision on a prespecified
target while Jev correctly answers the requests actually submitted
to it. Correct request-level reasoning and correct task-level
decision making are therefore evaluated separately.

The adversary knows the policy, adapter interface, and decision
rule, but not the reference labels or hidden auditor verdicts.
Each study specifies its permitted change and search budget.
We measure the effect on the prespecified target cases and report
available clean or non-target controls. We do not claim a uniform
selectivity rate across all three pillars.

\begin{table*}[!t]
\centering
\caption{Primary paired-study design. The source, policy, selector,
and authentic API record remain protected in every comparison.}
\label{tab:study-design}
\renewcommand{\arraystretch}{1.18}
\begin{tabular}{@{}p{0.11\textwidth}p{0.25\textwidth}p{0.35\textwidth}p{0.18\textwidth}@{}}
\hline
Pillar & Case unit & Changed adapter output & Bound \\
\hline
State & Eligibility records; six prespecified denial targets among 12 originals & One extracted expiry advances while the source stays fixed & At most five days and five proposals per target \\
Question & 12 authored domains with a failed policy prerequisite & A permission question becomes a denial-form question; state and selector stay fixed & One question output \\
Output & 12 records eligible earlier but ineligible after an update & The interpreter selects an authentic earlier answer for the current decision & One answer association; no response-byte change \\
\hline
\end{tabular}
\end{table*}
\section{Attack}
The pipeline has three interfaces at which the meaning of a decision
is carried forward. The structured state $s$ asserts which facts were
extracted from source $x$; the question $q$ specifies what Jev is asked
to judge under policy $\pi$; and the interpreted evidence $e$ tells
the final selector how to use Jev's authentic response $r$. A correct
response to the submitted request does not by itself verify any of
these connections to the original task. Figure~\ref{fig:interface-attacks}
shows the three points at which one compromised adapter can change
that connection. We examine each point separately below.
\subsection{Pillar I: State Manipulation}

The adversary controls the structured state emitted by the extraction
adapter, replacing $s$ with $\widetilde{s}$. An intervention can alter a
permitted field while preserving the required schema. Subsequent
question generation and reasoning operate on the altered state.
The integrity question is whether the state still represents the
decision-relevant facts in the protected raw input.
For example, a credential document can state an expiry that precedes
the requested day, while the adapter advances only the extracted
expiry. Jev can correctly assess the altered state as eligible even
though the source record supports denial. The attack changes the
evidence on which the judgment is conditioned, not the judgment
algorithm.

\subsection{Pillar II: Question Manipulation}
The adversary controls the question-generation adapter, replacing $q$
with $\widetilde{q}$ while leaving the extracted state unchanged.
An intervention can change the question's scope or omit a required
condition within the declared budget. The reasoning service answers
the question it actually receives. The integrity question is whether
that question still expresses the intended policy and decision task.
The most consequential change in our study preserves the policy's
conditions but reverses the meaning of the answer. If a protected
selector treats \emph{yes} as approval, an adapter can instead ask
whether approval should be withheld. A correct \emph{yes} to that
question then authorizes the opposite action. A check for policy terms
alone can miss this reversal because the requirements still appear in
the question.

\subsection{Pillar III: Output Manipulation}
The adversary controls the downstream interpretation adapter,
replacing $e$ with $\widetilde{e}$. Here, ``output'' means the evidence
representation derived from the reasoning response, not the response
returned by Jev itself. An intervention can change an allowed derived
value, omit evidence, or change an evidence association, subject to
the specified interface. The authentic response remains intact, and
the protected decision rule consumes the altered evidence.
In our versioned-record case, the adapter selects an authentic Jev
answer to an earlier, eligible state after a suspension or other
update makes the current state ineligible. The selected answer remains
correct for its own request; the violation is its use as evidence for
a later decision. This separates response authenticity from response
relevance.

\subsection{Attacker Objective and Audit Visibility}
The adversary seeks a prespecified incorrect decision on target inputs
while preserving reasoning correctness on the actual API requests.
Selectivity additionally requires measuring effects on non-target
inputs. For a prespecified target decision $y_{\mathrm{tar}}$, a counted
interface-integrity success requires
\begin{align}
 y &= y^*(x,\pi), \nonumber\\
 y' &= y_{\mathrm{tar}} \ne y^*(x,\pi), \label{eq:success}\\
 C(s,q,r) &= C(s',q',r')=1. \nonumber
\end{align}
The final condition distinguishes a wrong decision caused by an
adapter boundary from an ordinary error in Jev's answer to the
request it actually received.
Tildes denote the attacked execution, including any legitimate
downstream consequences of the selected intervention.

An auditor receives a specified projection $O(\tau)$ of the execution
trace $\tau$. Audit acceptance is an additional success condition,
reported separately for each observation setting. The settings are:
(i) the submitted state, question, and authentic response;
(ii) that exchange plus the final decision and decision rule; and
(iii) the full trace, including the original input, intended policy,
and interpreted evidence. The auditor cannot access hidden attack
labels. Apparent legitimacy is measured as acceptance under a stated
auditor and observation setting; it is not assumed from valid syntax.

For output-only interventions, clean and attacked paths can share the
same recorded API exchange. State and question interventions generally
change the exchange; they do not imply an unchanged service response.

\subsection{Scope}
Direct attacks on Jev, model training or parameter modification,
forged API responses, compromise of the final decision rule, and
simultaneous compromise of multiple adapters are outside scope.
The adapter boundary is imposed by the experimental interface rather
than enforced through process isolation. Our results therefore describe
the consequences of bounded, single-adapter control, not a mechanism
for obtaining or containing a compromise in a deployed system.
\begin{table}[t]
\centering
\caption{Technical stack for the primary and transfer studies.}
\label{tab:model-roles}
\small
\renewcommand{\arraystretch}{1.12}
\begin{tabular}{@{}p{0.41\columnwidth}p{0.49\columnwidth}@{}}
\hline
Component & Role and task \\
\hline
\texttt{jev-1.13.0} & Judgments and checks: all pillars \\
DeepSeek Flash & Question-equivalence check \\
GPT-4.1 mini (2025-04-14) & Question check; transfer labels \\
Fixed code & Adapter substitutions; selector \\
\hline
\end{tabular}
\end{table}

\section{Evaluation Design}
\label{sec:evaluation}
Our evaluation isolates the three intervention points in
Fig.~\ref{fig:interface-attacks}. Each study begins with a protected
source, policy, and reference decision. We record a clean execution,
then replace the output of exactly one adapter and let the remaining
pipeline operate on that value. The resulting primed trace tests
whether $y'$ differs from the reference decision while $r'$ correctly
answers the submitted $(s',q')$. Table~\ref{tab:study-design} gives
the case unit and intervention for each pillar. These are separate
studies, so their denominators are not pooled.

\subsection{Measurements and controls}
For each pair we classify the clean task decision, Jev's local
correctness on the request actually submitted, and the attacked
final decision. A locally incorrect Jev answer remains in the case
denominator but is not counted as an interface-integrity success.
We separately record whether a boundary check accepts the altered
trace and whether it permits legitimate clean work. Those defense
comparisons are reported in Section~\ref{sec:defense}; Section
\ref{sec:results} first establishes the decision effect.

The eligibility study also includes paraphrases and renewals for
clean-utility assessment. They are related representations, not
independent cases. A separate cross-pillar set applies the three
interventions to the same 36 actions under six rule structures.
Its labels received blinded-by-packet AI review, not human
adjudication. Five exact-request rounds on the original cases then
measure response stability with fixed attack choices. Repeated
calls do not increase task $N$. All reported Jev calls use
\texttt{jev-1.13.0} and a 0.5 decision threshold. The authored,
flip-capable tasks test mechanisms under bounded adapter control,
not the prevalence of compromise in deployed systems.

\section{Results}
\label{sec:results}
Table~\ref{tab:primary-outcomes} keeps local answer correctness
beside the resulting task decision. The clean state count includes
related representations, whereas its attack count concerns only
the prespecified targets. The question study retains a locally
incorrect attacked answer in its denominator. The output study
compares a current decision with an answer authentic to an earlier
request.

\begin{table*}[h]
\centering
\caption{Primary outcomes by pillar. Local correctness concerns the
request Jev actually received; a wrong final decision concerns the
protected task. Counts belong to different synthetic studies.}
\label{tab:primary-outcomes}
\renewcommand{\arraystretch}{1.18}
\begin{tabular}{@{}p{0.11\textwidth}p{0.29\textwidth}p{0.25\textwidth}p{0.24\textwidth}@{}}
\hline
Pillar & Clean task baseline & Attacked local correctness & Wrong final decision \\
\hline
State & 36/36 related representations correct before verification & 6/6 selected target requests correct & 6/6 targets granted \\
Question & 12/12 clean current-state requests correct & 11/12 denial-form requests correct & 11/12 wrong approvals from locally correct answers \\
Output & 12/12 current and 12/12 archived requests correct & 12/12 selected archived answers correct for their own requests & 12/12 wrong current grants \\
\hline
\end{tabular}
\end{table*}

\subsection{Three ways to lose task meaning}
The state row of Table~\ref{tab:primary-outcomes} is conditional on
a bounded search over small expiry changes. Jev correctly judged
the altered states, but those states no longer reflected the source.
The result shows why a well-formed request cannot establish source
fidelity; it is not the success rate of an arbitrary expiry edit.
The clean representations establish a baseline for this authored
record family without forming a larger independent sample.

In the question row, a positive answer to a denial-form proposition
was incompatible with code that interprets \emph{yes} as permission.
One attacked question received an incorrect Jev answer and is not
counted as an interface success. Keeping it visible separates an
ordinary model error from a correct judgment applied to the wrong
action. The behavior of the semantic and explicit question checks
is compared in Section~\ref{sec:defense}.

The output row isolates a different failure: the selected answer
was authentic and correct for its earlier request, yet irrelevant
to the current record. The result establishes the decision
consequence of stale evidence. Table~\ref{tab:defense-primary}
then shows which observers had enough context to detect its use.

\subsection{Breadth and response stability}
Table~\ref{tab:breadth-repeat} separates two follow-ups. The
cross-pillar set tests whether the three mechanisms appear under
other Boolean rule structures. The exact-request rounds test whether
the original outcomes depend on one Jev response. The transfer
cases were deliberately flip-capable, and the rounds reuse fixed
cases and attacks; neither column estimates real-world prevalence.

\begin{table}[t]
\centering
\caption{Non-pooled follow-ups. Transfer uses the same 36 actions
for three separate interventions; repeat statistics summarize five
exact-request rounds on the original cases.}
\label{tab:breadth-repeat}
\renewcommand{\arraystretch}{1.18}
\begin{tabular}{@{}p{0.18\columnwidth}p{0.27\columnwidth}p{0.39\columnwidth}@{}}
\hline
Pillar & Transfer: wrong decisions & Repeats: mean $\pm$ sample SD \\
\hline
State & 36/36 & 100\% $\pm$ 0 pp \\
Question & 36/36 & 95.0\% $\pm$ 4.6 pp \\
Output & 36/36 & 100\% $\pm$ 0 pp \\
\hline
\end{tabular}
\end{table}

In Table~\ref{tab:breadth-repeat}, the three transfer interventions
are counterfactuals on the same actions, not independent task
samples. Jev answered their submitted clean, earlier-version,
and denial-form questions correctly. The selected state and output
outcomes were stable across repeats; question variation came from
one answer near the decision threshold. A semantic checker rejected
the transfer set's direct denial wording despite accepting the
primary study's different denial wording. This contrast limits any
general claim about semantic checking; Section~\ref{sec:defense}
examines the boundary checks directly.

\section{Defending the Decision Boundary}
\label{sec:defense}
The three attacks exploit different missing relationships, so a check
of Jev's answer alone cannot address all of them. Jev is trusted here
as an uncompromised service, not as an infallible oracle. Its judgment
can be correct for the submitted request while the final decision is
wrong; it can also make an ordinary reasoning error on that request.
The defenses below address the first problem by preserving the
context needed to use an answer. They do not establish that every
answer Jev gives is correct.

Table~\ref{tab:defense-primary} matches each primary attack to the
checks actually evaluated. ``Wrong decisions allowed'' counts attacked
paths that both produce a wrong final decision and pass the named
check; it is not a detector's raw acceptance rate. Clean acceptance
uses the corresponding legitimate inputs, which differ by study.
The question clean control and corrected output full-trace audit were
specified after defects in their initial versions were found, so
their clean results are not independent held-out confirmations.

\begin{table*}[t]
\centering
\caption{Checks evaluated on the three primary studies. A dash means
that clean acceptance was not measured for that comparison. Clean
state inputs include related representations, while question and
output clean inputs are their respective study cases.}
\label{tab:defense-primary}
\renewcommand{\arraystretch}{1.18}
\begin{tabular}{@{}p{0.11\textwidth}p{0.39\textwidth}cc@{}}
\hline
Pillar & Check or audit view & Wrong decisions allowed & Clean accepted \\
\hline
State & Canonical question and direct output use, without source check & 6/6 & --- \\
State & Combined check including source--state consistency & 0/6 & 33/36 \\
\hline
Question & Semantic policy-equivalence check & 11/12 & 10/12 \\
Question & Explicit \emph{yes}-to-approval check & 0/12 & 12/12 \\
\hline
Output & Response-byte authenticity & 12/12 & 12/12 \\
Output & Auditor shown only the selected exchange & 12/12 & 12/12 \\
Output & Exact binding to the current request & 0/12 & 12/12 \\
Output & Corrected full-trace audit & 0/12 & 12/12 \\
\hline
\end{tabular}
\end{table*}

\subsection{Bind state to the source}
The state attack substitutes a plausible extracted fact for a fact
in the protected document. Schema validation cannot detect this:
both the clean and altered states have the same form. The relevant
check must compare decision-bearing state fields with the source and
its amendments before the state is submitted or used for approval.
The source-aware check rejected the selected state interventions in
our study, but it also rejected faithful clean representations, as
Table~\ref{tab:defense-primary} shows. A rejected state should therefore
trigger review or renewed extraction rather than be counted as a
correct denial. Source comparison is only as strong as the protected
record and the check's ability to interpret it.

\subsection{Bind the question to the action}
The question attack leaves the policy conditions visible but changes
what a positive answer means. A policy-equivalence check can accept
that wording while missing its operational polarity. The selector
needs an explicit contract between the proposition sent to Jev and
the action taken for each answer. In our fixed \emph{yes}-to-approval
setting, the answer-polarity check rejected the denial-form questions
that the semantic check accepted. A canonical permission question is
another simple control when the application does not need arbitrary
question generation. Neither measure corrects a wrong Jev answer;
it ensures that an answer is applied to the action it was asked about.

\subsection{Bind evidence to the current request}
The output attack reuses an authentic answer after its underlying
record changes. Byte authenticity verifies origin, but not which
decision the answer supports~\cite{lowe1997authentication}. The interpreter must associate each
response with the state, question, and record version of its own
request, then compare that identity with the current decision before
the selector consumes it. This binding rejected the stale selections
in Table~\ref{tab:defense-primary}. An auditor shown only the archived
exchange cannot make the comparison because the later update is
absent from its view; a full-trace auditor can, provided its own
question and verdict mapping are correct.

These checks protect different links in the same chain. An output
binding cannot recover a fact already changed upstream, and a
source-aware state check does not establish that \emph{yes} still
authorizes the intended action. In the tested cases, the appropriate
boundary check stopped each corresponding intervention. This is a
limited defense result for bounded, single-adapter compromise, not
evidence that the checks are complete against other manipulations or
that they eliminate independent errors by Jev.

The separate cross-pillar set gives a further comparison across six
rule structures (Table~\ref{tab:defense-transfer}). Its explicit
checks again rejected the constructed attacks. Unlike the primary
studies, this set did not measure the corresponding clean-acceptance
rates, so it cannot establish their utility on new rule structures.
The semantic question checker also rejected the altered wording in
this set, in contrast to its acceptance of the earlier denial-form
wording. This difference limits any claim that semantic checking
always fails at the question boundary.

\begin{table}[t]
\centering
\caption{Defense transfer on a separate set of 36 deliberately
flip-capable cases. Each pillar is a separate intervention on the
same cases; clean acceptance was not measured here.}
\label{tab:defense-transfer}
\renewcommand{\arraystretch}{1.18}
\begin{tabular}{@{}p{0.21\columnwidth}p{0.48\columnwidth}c@{}}
\hline
Pillar & Explicit check & Attacks accepted \\
\hline
State & Current-state equality & 0/36 \\
Question & Answer-polarity audit & 0/36 \\
Output & Current-request binding & 0/36 \\
\hline
\end{tabular}
\end{table}

\FloatBarrier
\section{Discussion and Future Directions}
\label{sec:discussion}
The primary studies isolate three decision-integrity failures, not
the frequency with which real adapters become compromised. The
follow-up runs suggest harder tests of the same boundaries. For
state, a source check must preserve relationships among entities,
permissions, and amendments while permitting faithful extractions.
For questions, agreement on the observed state may be insufficient
if a generated question changes the policy's answer on a relevant
counterfactual state. For output, provenance of an authentic answer
may be necessary but insufficient when an application must combine
several current judgments. These exploratory observations motivate
held-out, independently reviewed cases and comparisons at matched
clean acceptance and cost; they are not additional cases in the
primary result tables.

\section{Conclusion}
A trusted typed-judgment service does not by itself guarantee a
correct application decision. In our Jev-based pilot, bounded changes
at the state, question, and output interfaces produced wrong final
decisions even when the relevant Jev answers were authentic and
locally correct. The tested checks stopped their corresponding
interventions by binding extracted facts to the source, answer
meaning to the intended action, and evidence to the current request.
The result is a three-boundary account of where decision meaning can
be lost, together with a limited empirical test of how that meaning
can be preserved.

\clearpage 
\bibliographystyle{IEEEtran}
\bibliography{references}
\end{document}